# Low Actuation Voltage Totally Free Flexible RF MEMS Switch With Antistiction System

S. Touati[1], N. Lorphelin[1], A. Kanciurzewski[1], R. Robin[1,2], A.-S. Rollier[1], O. Millet[1], K. Segueni[1]
[1]DELFMEMS SAS, 5 rue Héloïse, Haute Borne, 59650 Villeneuve d'Ascq - France
[2]Institut d'Electronique de Microélectronique et de Nanotechnologie, IEMN, Villeneuve d'Acsq, France

*Abstract*- **This paper concerns a new design of RF MEMS switch combined with an innovative process which enable low actuation voltage (<5V) and avoid stiction. First, the structure described with principal design issues, the corresponding anti-stiction system is presented and FEM simulations are done. Then, a short description of the process flow based on two non polymer sacrificial layers. Finally, RF measurements are presented and preliminary experimental protocol and results of anti-stiction validation is detailed. Resulting RF performances are -30dB of isolation and -0.45dB of insertion loss at 10 GHz.**

## I. INTRODUCTION

A frequent cause of failure in RF MEMS (Radio Frequency MicroElectroMechanical Systems) switches is stiction phenomenon due to capillarity adhesion, dielectric charging and contact adhesion [1],[2]. The stiction occurs especially with small gaps, low stiffness and large/smooth areas.

The dielectric charging that could shift the actuation voltage and imply a permanent collapse state is decreased by adapted methods of deposition, the choice of low default dielectric layer or a no-contact at the actuated state between electrodes and moving parts. Contact adhesion is a more critical aspect to solve.

Due to low contact forces reachable by MEMS structures (< 1mN), a soft metal is needed to obtain low contact resistance (1 ohms). Gold with a stable contact resistance inferior to 1 ohm for a contact force of 100µN is the standard candidate for contact metal for MEMS switches [3],[4]. Nevertheless the high adhesion force of gold compared to mechanical restoring forces implies a failure mode of switches in the on state. A system to improve gold contact switches reliability could be done by added forces to counter metal surface adhesion. A solution based on a simply supported free structure meets the requirement for an effective solution [5]. The MEMS switch is supported by two pillars, on both side two different electrodes are placed.

For a series switch, the two internal actuation electrodes induce negative out of plane deflection to improve contact forces at the on-state. Two external electrodes induce a positive out of plane deflection to improve isolation at the off-state. This topology allows obtaining an active restoring force for the two functioning states of the membrane with a complexity level very limited (one level of actuation electrodes) compared to other solutions [6], Fig. 1.

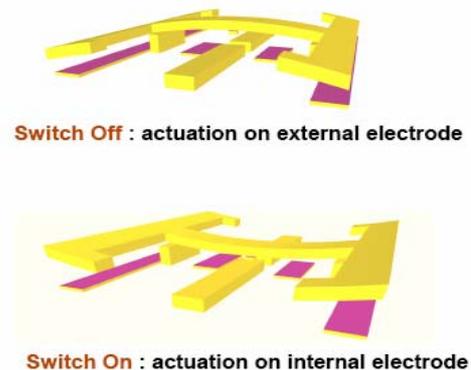

Fig. 1. 3D illustration of the two states of the switch.

Three different mechanical states can be distinguished: rest state (Fig. 2(a)) that corresponds to the mechanical states after the releasing of the structure. The forced off state performed by actuation on external electrodes (Fig. 2(b)) and the forced on state performed by the actuation on internal electrodes (Fig. 2(c)).

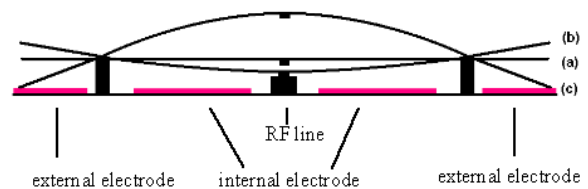

Fig. 2. (a) Rest, (b) on and (c) off states of the membrane.

This membrane allows to combine both low initial gap (<1µm) and large deflection (>2µm). Concerning the external actuation, the combination of lever effect and pillars position induces a large deflection (>2µm for initial gap <1µm) and generate high forces that counter stiction phenomenon (i.e. generated forces > stiction forces).

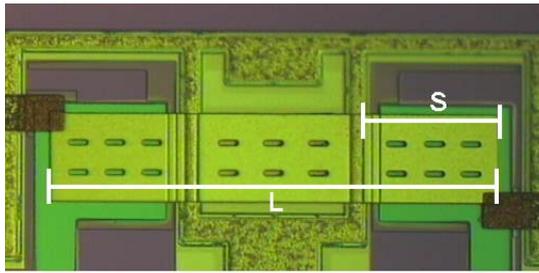

Fig. 3. Optical image of simplified membrane.

The ratio S/L between the length external to pillar and the total length of the membrane fixed the maximum deflection of the membrane at pull-in voltage, Fig. 3. For a 360μm length membrane and a gap of 1 μm, we have a deflection of 1.8 μm and a actuation voltage of 3.5 volts for a 0.1 ratio, Fig. 4.

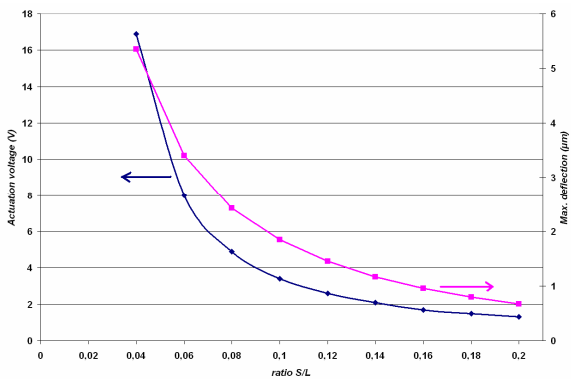

Fig. 4. Simulation of deflection and pull-in voltage versus S/L ratio.

This anti-stiction phenomenon and the property of mechanical structure (totally free membrane) increase intrinsic in-use (stiction, creep) and environmental reliability (temperature, vibrations).

## II. DESIGN

The membrane profile is changed to have an H-shape. The local increases of actuations area produce higher electrostatic forces. In the same way, the gap has been reduced to a very low value (<1μm) [7], [8]. FEM simulations have been performed to exhibit comportment of the membrane before and after pull-in for the 360μm length membrane. Due to the lever effect, displacements before pull-in are superior to the classical limit of clamped beams. The maximum deflection is 0.75μm for a gap of 1 μm.

The comportment after pull-in is interesting because the deflection remains constant by increase of the actuation voltage. The collapse area on external electrodes remains stable for moderate "over pull-in" actuation voltage, Fig. 5.

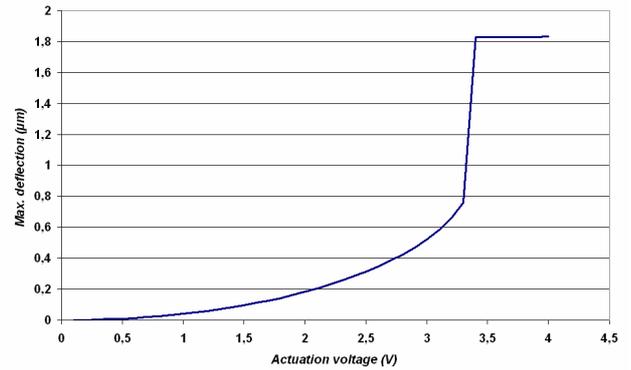

Fig. 5. Simulation of deflection versus actuation voltage.

The increase of membrane width in the H part implies a zipping effect that decreases the efficiency of actuation concerning maximum deflection. To avoid this phenomenon 4 supports points, named wings, are added. Fig. 6, illustrate the deflection without and with wings, maximum deflection at pull-in voltage is multiply by 2 (0.94 versus 2 μm) with wings addition.

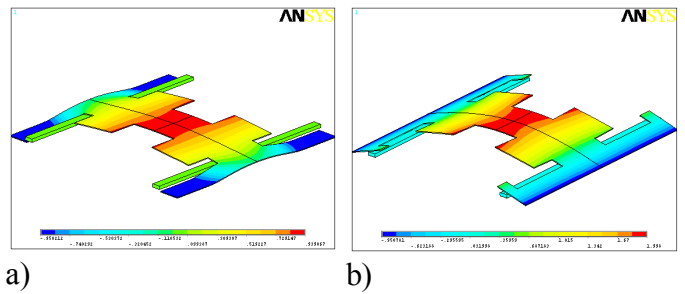

a)          b)

Fig. 6. FEM simulations of external actuation without a) and with b) wings.

To avoid sliding of the membrane, mechanical stop units (MSU) are fabricated by auto-alignment, Fig. 7.

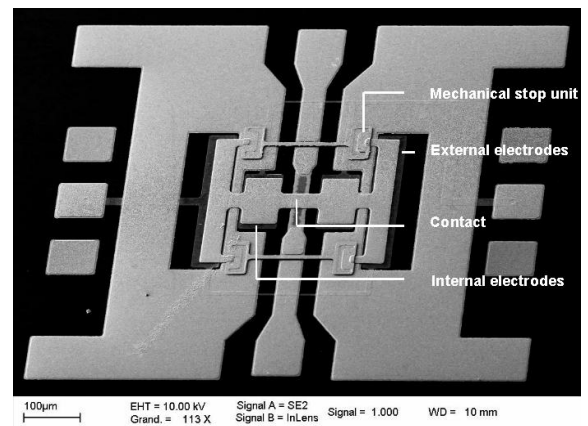

Fig. 7. SEM view of the MEMS structure.

 

ANSYS™ simulations have been done to compare restoring forces for a cantilever, a fixed-fixed membrane and the membrane, Tab. 1. For each type of structure, a model has been designed and pressure forces have been applied to calculate the minimum pressure which induces contact between the free part and the fixed internal electrodes. Results show that the membrane restoring force is higher than the cantilever one but lower than the fixed-fixed beam one. So without external actuation the membrane is more sensitive to stiction than a fixed-fixed beam.

TABLE I
MINIMUM PRESSURE INDUCING CONTACT FOR EACH TYPE OF COMPONENT

| Type of MEMS | Minimum pressure inducing contact (MPa) |
|---|---|
| Cantilever beam | 1e-4 |
| Clamped-clamped beam | 2e-3 |
| Membrane without external actuation | 1e-3 |
| Membrane with external actuation | 5e-3 |

Concerning the external actuation, the design has been optimized to perform large deflection and a stable state. By the way, a minimum collapse area has been done to avoid stiction concerning off-state.

The increase of H parts area and the addition of balancing wings allow improving the efficiency and the level of generated forces to counter possible on state stiction, Fig.8.

In contrast with classical MEMS switches, an active electrostatic restoring force is performed to improve the reliability of high adhesion contact metal switches.

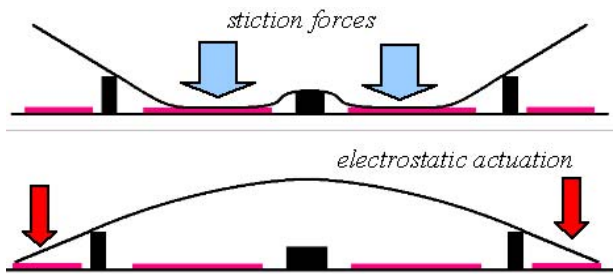

Fig. 8. Stuck membrane and unsticking due to external actuation.

The anti-stiction system has been validated by numerical simulation and confirmed by experiments. Electrostatic actuation of a 360μm length, 250μm width and 1μm thick membrane has been simulated. With a 0.7μm gap, internal pull-in occurs around 3.5V and pull-out at 1.4V due to mechanical restoring forces. To simulate dielectric charging, a maximum of 2V has been applied on internal electrodes. Simulation exhibits a un-stiction of the structure at 5V on external electrodes (Fig. 9). This actuation acts as an electrostatic restoring system.

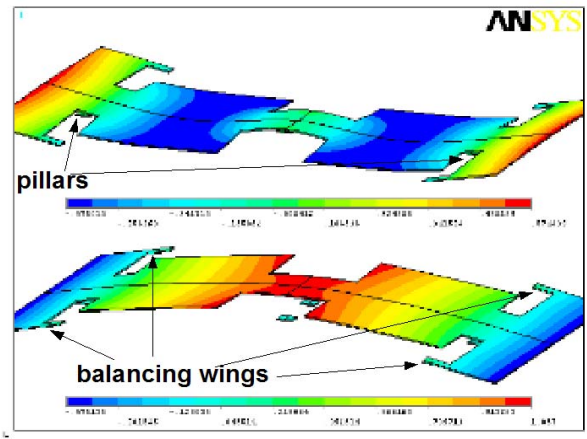

Fig. 9. Illustration of FEM simulations of sticked and un-sticked states.

Fig. 10 shows the membrane profile at different states: down state, stuck state and unstuck state. The mechanical restoring forces are not sufficient to totally un-stick the membrane if the voltage is higher than 1.5V. Then, the external actuation is necessary to counter stiction.

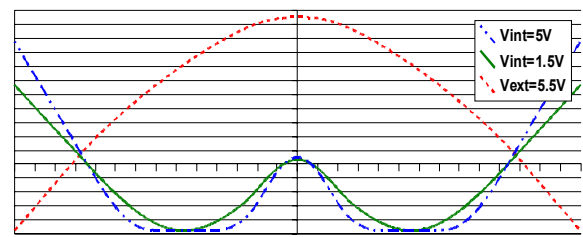

Fig. 10. Membrane profile for internal pull-in, stuck and unstuck state.

### III. FABRICATION PROCESS

The fabrication process is based on two sacrificial layers and 9 masks level, Fig. 11. The challenge of the process flow is to obtain a free membrane fixed on a functioning position after releasing step.

On silicon wafer, actuation electrodes are patterned by gold sputtering and wet etching. The electrode insulation is obtained thanks to 200nm thick PECVD (Plasma Enhanced Chemical Vapor Deposition) silicon nitride and RIE (Reactive Ion Etching) plasma etching, Fig. 11-a.

A gold layer is then patterned to form the seed layer for RF lines growing, Fig. 11-b. A Cr layer is sputtered on the wafer as sticking layer for the PECVD silicon dioxide, this layer ensure electrical connection for further electroplating. The gold electroplating of RF lines and pillars is performed with the silicon dioxide mold etched by RIE plasma and chromium by wet etching, Fig. 11-c. The electrostatic gap is also defined by the mold thickness (1st sacrificial layer).

A sputtered Cr layer is used as 2[nd] sacrificial layer to obtain free membrane, Fig. 11-d. Ohmic contacts composed of gold and silicon nitride are patterned respectively by wet and dry etching.

The 1μm metallic membrane is performed by optimized electroplating and wet etching to obtain very low stress gradient. The final part of sacrificial layer is still a sputtered chromium layer, Fig. 11-e. The mechanical stop units are auto-aligned and performed by gold electrolysis and local aperture of chromium for anchoring. The releasing step is divided on two steps. The first one consists on the silicon dioxide etching by an hydrofluoric base solution. At the step membrane remains fixed by the chromium layer.

The second one consists on the chromium removing by a commercial selective etchant, Fig. 11-f. To suppress stiction problem, a $CO_2$ supercritical drying is finally performed.

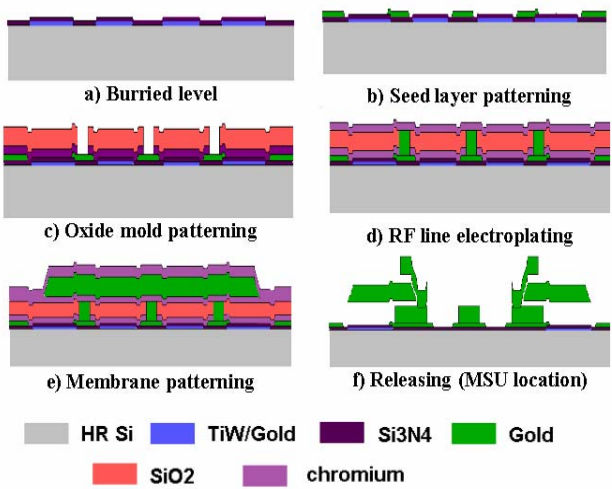

Fig. 11. Description of process flow.

This process presents several advantages:
- A low temperature process < 300°C.
- No polymer as sacrificial layer.
- Two sacrificial layers for electrostatic gap and mechanical stop units auto-alignment.

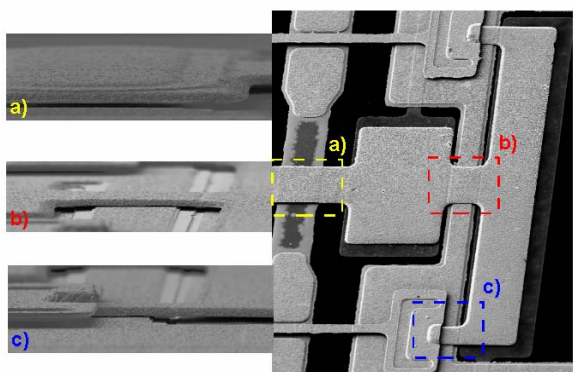

Fig. 12. SEM views of released structure, a) contact area, b) pillar area and c) wings and mechanical stop unit area

## IV. TESTING

### A. Actuation cycles

The devices have been tested with a probe station in the air environment just after releasing, Fig. 12. Pull-in measurements have been performed asynchronously on internal and external electrodes. For the both states, the voltage is less than 7V by considering dispersion on capacitive measurements done by impedencemeter. In parallel, video captures have been performed. Fig. 13 illustrates the 3 states of the membrane for a shunt switch.

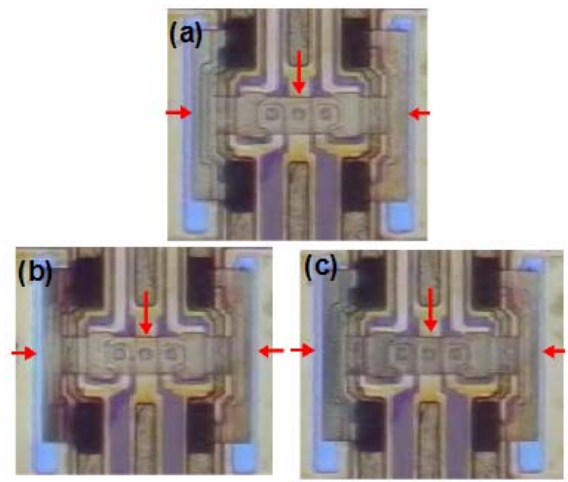

Fig. 13. Optical images of electrostatic actuation, (a) rest state, (b) on-state, (c) off-state.

### B. RF testing

S-parameters of the RF MEMS switches have been measured on a HP network with a probing system. The devices have been tested on a frequency range of 0.25-10GHz at up state and down state. For the operating frequency of 10GHz the measured isolation is better than -30dB and the insertion loss is inferior to -0.45dB HR silicon substrate, Fig. 14.

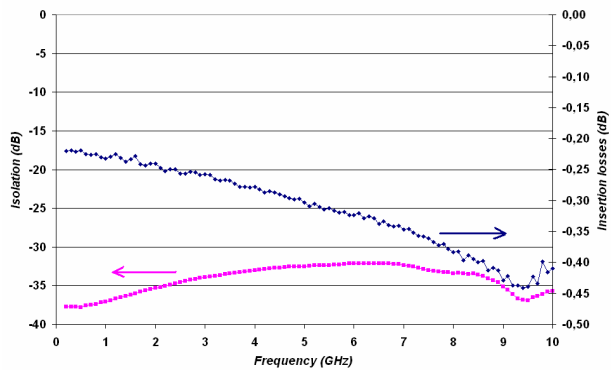

Fig. 14. S parameters of shunt ohmic switch.

*C. Anti-stiction validation*

To characterize stiction phenomenon high voltages (>15V) have been applied on internal electrodes during several minutes (≈15min). The membrane has been un-sticked with a maximum of 5V applied on external electrodes. All states have been characterized by resistive measurements between the membrane and RF line.

The anti-stiction phenomenon has also been validated on RF functioning on several devices. After actuating at 15V on internal electrodes during 10min, stiction due to dielectric charging has been observed. The on-stage remains at 0V. The application of 10V on external electrodes enables to un-stuck the membrane. The same RF performances are obtained. The un-stuck voltage difference is due to moisture of silicon nitride by long term exposition in air.

Next measurements will be done in environmental conditions to provide additional data on anti-stiction system versus actuation cycles.

V. CONCLUSION

A new design of RF MEMS switch has been presented based on a totally free flexible membrane fabricated with an innovative micro-machining process flow. This process allows to create a design which enables low actuation voltage (<5V) and a better reliability. The anti-stiction system has been validated by both finite element method simulation and characterizations. The MEMS structure provides the opportunity to increase the actuation cycle limitation of MEMS switches due to contact stiction with high surface adhesion metal.

**Acknowledgements**

The authors would like to thanks IEMN for the access to a high technological level clean room and MC2-technologies during the devices characterizations.